# Multipath Compensation Algorithm for TDMA-Based Ultrasonic Local Positioning Systems

T. Aguilera, F. J. Álvarez, D. Gualda, J. M. Villadangos, A. Hernández and J. Ureña.

*Abstract*—This work proposes a Multipath Compensation Algorithm (MCA) to enhance the performance of an Ultrasonic Local Positioning System (ULPS) under adverse multipath conditions. The proposed algorithm is based on the accurate estimation of the environment impulse response from which the corresponding Line of Sight (LoS) for each channel is obtained. Experimental results in two different environments and with different conditions have been conducted in order to evaluate the performance of this proposal. In both environments, results confirm the expected improvements, even under severe multipath conditions where positioning errors have been reduced from 44 cm to 9 cm for the 95% of the measurements.

## I. INTRODUCTION

THE availability of the different technological approaches appeared in recent years has enabled the estimations of position for an object and/or person with enough accuracy in indoor environments. This positioning information is key to provide Location-based Services (LBS) in a wide range of application fields, from industry or entertainment to elderly monitoring and healthcare.

Global Positioning System (GPS) provides an accurate and robust user location anywhere on Earth, whether there is a direct Line-of-Sight (LoS) with four or more GPS satellites. Nevertheless, GPS is not suitable to estimate indoor positions due to the attenuation and multipath effects produced by obstacles and walls on the GPS signals. To extend the user's location to indoor environments different sensory signals, such as Radio-Frequency (RF) [1], [2], [3] or InfraRed (IR) [4], can be used by the beacons, which behave as an indoor satellites, to compose the transmissions acquired by the receiver in order to carry out the estimation of its own position. These approaches achieve similar accuracies and performance to those provided by GPS (few meters) [5], [6], [7]. Likewise, the ultrasonic technology has also contributed with interesting achievements [8], [9], [10] and significant advantages, such as its relative low cost and its centimeter accuracy. On the contrary, it also presents some limitations, such as the reduced coverage

Manuscript received June 20th, 2017
This work was funded by the Spanish Ministry of Economy and Competitiveness and the European Regional Development Fund (ERDF), under projects TARSIUS (ref. TIN2015-71564-C4-1/4-R), SOC-PLC (ref. TEC2015-64835-C3-2-R), and by the Regional Government of Extremadura and ERDF under project "Consolidation of Research Groups" (GR15167)

T. Aguilera and F. J. Álvarez are with the Department of Electrical Engineering, Electronics and Automation, University of Extremadura, E-06006, Badajoz, Spain. E-mail: {teoaguibe, fafranco}@unex.es.
D. Gualda, J. M. Villadangos, A. Hernández and J. Ureña are with the Department of Electronics, University of Alcalá, E-28805, Alcalá de Henares, Spain. E-mail: {david.gualda, villa, alvaro, urena}@depeca.uah.es.

area or the multipath and near-far effects, which restrict its implementation under certain circumstances and affect the final system performance [11].

Target location is a key aspect closely related to any Local Positioning System (LPS), no matter the technology used in its development. Previous works have already surveyed this aspect [5], [12], [13], deeply describing algorithms and methods proposed to determine the location of a target object, person or mobile robot, often for assisting or guiding purposes. Different approaches can be found depending on some general design parameters, such as the privacy orientation of the system, the positioning algorithm implemented, the number of target objects to be located simultaneously, or the characteristics of the environment to be explored. In this sense, a detailed review of ultrasonic local positioning systems focused on target location can be found in [14].

Generally speaking, an Ultrasonic Local Positioning System (ULPS) is composed of a set of at least three beacons strategically located to provide a suitable coverage throughout the positioning area [8]. Furthermore, their transmissions are normally encoded and concurrent to avoid any synchronization link between beacons (emitters) and a possible receiver [15], [16], [17], [18]. The use of different types of sequences [19] in the transmission encoding, such as pseudo-random codes [20] or complementary sets of sequences [21], allows precise and robust times-of-arrival (ToA) to be estimated, even under weak signal receptions or noisy environments, thanks to their suitable correlation properties. Unfortunately, the use of encoded emissions leads to longer transmissions, which may increase the multipath effect in those environments where walls, corners or other obstacles generate undesired path detections in the receiver. Some previous works have already dealt with this issue, either trying to avoid multipath-affected environments if possible [8] [10], or openly admitting a clear performance deterioration under these difficult conditions [22], [23], [24], [25].

In indoor environments with a severe multipath effect, the secondary paths may be numerous and closely spaced over time. In addition, multipath dynamically changes indoors, depending on the environmental conditions; i.e, the secondary paths may vary, either disappearing or appearing new ones as time goes by. Furthermore, the LoS is weaker than some secondary paths, or even it may not be available at all. In both cases, the firstly arrived secondary path is wrongly identified as the LoS. In that situation, conventional algorithms based on the maximum value of the correlation output clearly fail. Nevertheless, the LoS signal is not as variable as secondary paths are over time. This feature can be used to correctly identify the LoS signal, thus achieving a better ToA estimation.

Multipath effect has been widely studied, especially for outdoor GPS applications [26]. Nonetheless, some recent works focused on multipath cancelation strategies for Acoustic Local Positioning Systems (ALPS) have been proposed. For example, in [27] a Time Division Multiple Access (TDMA)-based system is implemented to locate a mobile smart device by using the emission of pseudo-noise cyclic sequences. This system benefits from the a priori known features of the cyclic emissions to discard the arrival of reflected echoes by applying a time-sliding window. It is also worth noting the work proposed in [28], where an iterative Peak Matching algorithm is designed to cancel the multipath effects that appear in the auto-calibration process of a wireless network of acoustic beacons.

On the other hand, a different approach against multipath has already been proposed in [29], where a Code Division Multiple Access (CDMA)-based ALPS is implemented to provide indoor positioning for portable devices that acquire the encoded signals simultaneously emitted by a set of fixed beacons. The afore-mentioned ALPS is mainly oriented to smart devices (smartphones and tablets), and it consists of four beacons distributed in front of the user. Their emissions are in the audible range (16 kHz carrier frequency), and they are based on 63-bit Kasami sequences, applying a CDMA scheme. The core of the proposal is a Matching Pursuit algorithm [30], [31].

With regard to this, this work proposes the application of a multipath cancellation algorithm in an ULPS. Furthermore, the system uses a TDMA scheme and an encoding based on 255-bit Kasami sequences. The multipath cancellation algorithm has been applied in the correlation process in order to improve the determination of ToAs for the transmissions from every beacon under conditions characterized by a strong multipath. Experimental results have shown the feasibility of the proposal, achieving a significant improvement in the position estimation. A preliminary study have already been published in [32], whereas this work extends a thorough set of new experimental tests carried out in two different environments to verify the approach under severe multipath conditions, as well as it conducts a study of the proposed algorithm performance depending on the configuration of the corresponding design parameters.

The rest of the manuscript is organized as follows: Section II describes the ULPS used in this work; Section III details the proposed Multipath Cancellation Algorithm (MCA); Section IV shows some experimental results; and, finally, conclusions are discussed in Section V.

## II. GLOBAL SYSTEM OVERVIEW

### A. Emitter Module

The emitter module depicted in Fig. 1 a) has been developed at the Department of Electronics from the University of Alcalá [33]. This module is composed of five ultrasonic transducers Prowave 328ST160 [34], as the one shown in Fig. 1 b), whose frequency response is represented in Fig. 1 c). These transducers are distributed in a $0.7 \times 0.7$ m squared structure. The management of the ultrasonic beacons, together with the signal processing, has been carried out on a platform based on a LPC1768 microcontroller [35], specially designed for that purpose. This dedicated platform can be observed in Fig. 1 d), where its main blocks have been defined. Since it includes only one Digital-Analog Converter (DAC), a TDMA scheme is necessary to sequentially drive each beacon, thus transmitting every ultrasonic signal within a $\tau = 20$ ms time slot.

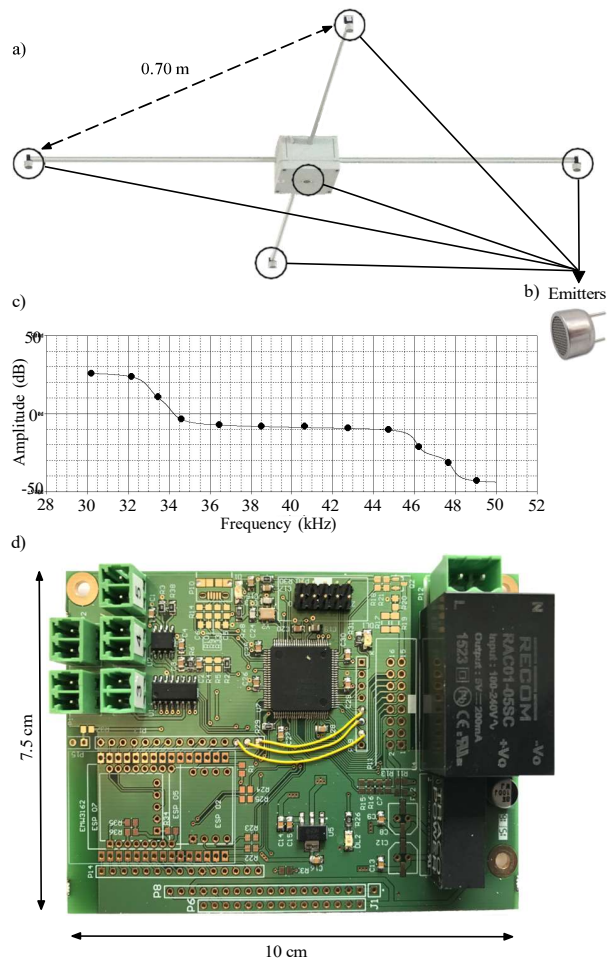

Figure 1. Ultrasonic local positioning system (a); transducer image (b); transducer frequency response (c); and microcontroller-based control unit (d).

The use of ultrasonic encoded transmission makes possible the identification of every emitter for the different ToAs measured at the receiver. Every beacon transmits a particular 255-bit Kasami code [36], which has been Binary Phase Shift Keying (BPSK) modulated, with a modulation symbol consisting of two periods from a 41.67 kHz sine carrier sampled at 500 kHz.

### B. Receiver Module

The receiver module, depicted in Fig. 2, is formed by a MEMS microphone SPU0414HR5H-SB [37] and a microcontroller STM32F103 [38], both integrated in an specific board designed for this system. This microcontroller captures the received signals at a 100 kHz sampling rate, and then it performs a conditioning stage (high-pass filtering and amplification).

The input buffer size is set to 10000 samples, constrained by the microcontroller's low memory capacity. Whether this buffer becomes full, data are sent via USB to a personal computer (PC), where the receiver's position is estimated.

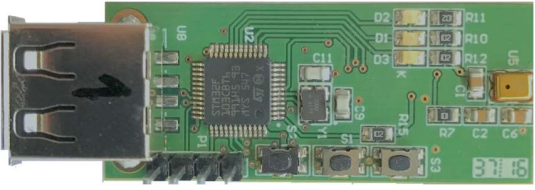

Figure 2. Receiver module based on a microphone SPU0414HR5H-SB and a microcontroller STM32F103.

Preliminary experiments carried out with the above described ULPS have already shown how important the multipath effect can be in terms of accuracy and reliability. It becomes even more relevant in certain environments where reflections on walls or obstacles strongly affect the correlation properties and, consequently, the ToA detection. This circumstance is graphically detailed in Fig. 3, where the zoomed correlation function for the 255-bit Kasami code transmitted by beacon no. 4 shows how the main correlation lobe cannot be distinguished from the sidelobes, thus making not possible to identify the ToA for the direct path transmission. Note that this signal corresponds to the transmission of a 255-bit Kasami sequence, BPSK modulated.

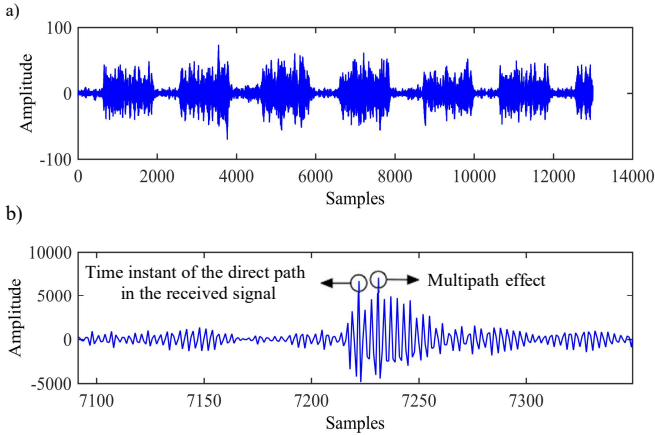

Figure 3. Example of a received signal (a); and zoomed correlation function for code no. 4 (b).

## III. Proposed Multipath Compensation Algorithm

Multipath is a propagation effect originated when the same acoustic signal reaches the receiver several times by following two or more different paths. This effect can be disruptive in the proximity of an obstacle or near the room walls, where multipath generates constructive and destructive (responsible for fading) interference and phase shifting of the received signal. The reflected signals interfere with the LoS transmissions, thus degrading the ideal correlation properties coming from the corresponding sequences, which implies a wrong ToA estimation. Nevertheless, considering that the time of occurrence for the first non-null coefficient in the impulse response represents the desired time-of-arrival (LoS-ToA), the wrong determination of the LoS-ToA produced by multipath can be avoided if an accurate estimation of the environment impulse response is obtained.

Assuming an environment with $I = 5$ channels, its corresponding received signal $r$ can be expressed as (1):

$$r = \sum_{i=1}^{I} S^i h^i + n \qquad (1)$$

where $S^i$ y $h^i$ are the Toeplitz Matrix (TM) of the emitted pattern and the impulse response of the $i$-th channel respectively; and $n$ is white Gaussian noise.

The operation of the MCA is detailed as follows:

1) At each iteration $j$ a new component of the impulse response for every channel $i$ is simultaneously estimated by the MCA. That is, for the $j$-th iteration and channel $i$, with $i = 1, \ldots, 5$, the algorithm calculates the impulse response components with amplitude $\hat{h}_j^i$, located at positions $l_j^i$, according to (2):

$$\hat{h}_j^i = \frac{c_{l_j^i}^{i\,T} r_j}{\|c_{l_j^i}^i\|^2} \qquad (2)$$

where $c_{l_j^i}^i$ is the $l$-th column of the TM for the $i$-th channel in the $j$-th iteration.

2) The MCA benefits from the TDMA scheme which provides a cyclical emission of signals with a constant interval $\tau$ between them and following an order previously established. With regard to this, the algorithm evaluates if the time difference between two consecutive impulse response locations, $l^i$ and $l^{i+1}$, are inside a timing window width $\tau \pm \delta$, where $\delta$ is the maximum difference time of arrivals for signals coming from different beacons.

3) If all locations $l^i$ are within their corresponding timing windows $[\tau \pm \delta]$ (3):

$$\forall i, \; \exists\, h_j^i \setminus l_j^{i+1} - l_j^i \in [\tau \pm \delta] \quad \text{with } i = 1,\ldots,5 \qquad (3)$$

the MCA selects the component with the highest amplitude $h_j^i$ in the set of channels $i$ for each iteration $j$:

$$a_j = arg \; \max_i \; h_j^i \qquad \text{with } i = 1,\ldots,5 \qquad (4)$$

where the $arg$ function means the value of $i$ that makes the expression $(max h_j^i)$. Then, the selected component $h_j^{aj}$ is stored and the remaining ones are discarded.

$$\hat{h}_j^{a_j}(l_j^{aj}) = h_j^{aj} \qquad (5)$$

On the contrary, if any location is outside its assigned timing window $[\tau \pm \delta]$ (6):

$$a_j = arg \; l_j^{i+1} - l_j^i \notin [\tau \pm \delta] \qquad \text{with } i = 1,\ldots,5 \qquad (6)$$

the corresponding spurious channel component is selected for next step in order to subtract its contribution from the received signal $r$. In this case, the MCA discards all the channel components for storage.

4) A new residue $r_{j+1}$ is calculated by using the impulse response from the selected component $h_j^{ajl}$. This residue will be used in the next iteration $j + 1$ to obtain the following five impulse response components $h_{j+1}^i$ (7).

$$r_{j+1} = r_j - h_j^{ajl} \, c_{l_j^a j}^{ajl} \qquad (7)$$

This process continues until a minimum number $M$ of components $h_j^i$ of the impulse response for each channel $i$ or a maximum number $J$ of iterations is achieved. These parameters can be modified if necessary.

5) After the procedure above described has reached the last iteration $j = J$, the LoS-ToA component must be chosen among all the previously calculated coefficients $h^{\hat{} i}$. For this, only those coefficients above a certain threshold will be considered as possible candidates, where this threshold is established as a fraction $\gamma$ of the highest component $h^{\hat{} i}$ in the channel impulse response. Therefore (8):

$$b^i = l^i \setminus h^{\hat{} i}(l^i) > \gamma \; \max(h^{\hat{} i}) \qquad (8)$$

Finally, the minimum value $\min(\mathbf{b}^i)$ is considered hereinafter as the LoS-ToA of the $i$-th channel (9):

$$\text{LoS-ToA}^i = \min(\mathbf{b}^i) \text{ with } i = 1, \ldots, 5 \qquad (9)$$

Both parameters, the fraction $\gamma$ and the minimum number $M$ of components in the channel impulse response are variables that determine the algorithm's performance. The general scheme of this method is shown in Fig. 4.

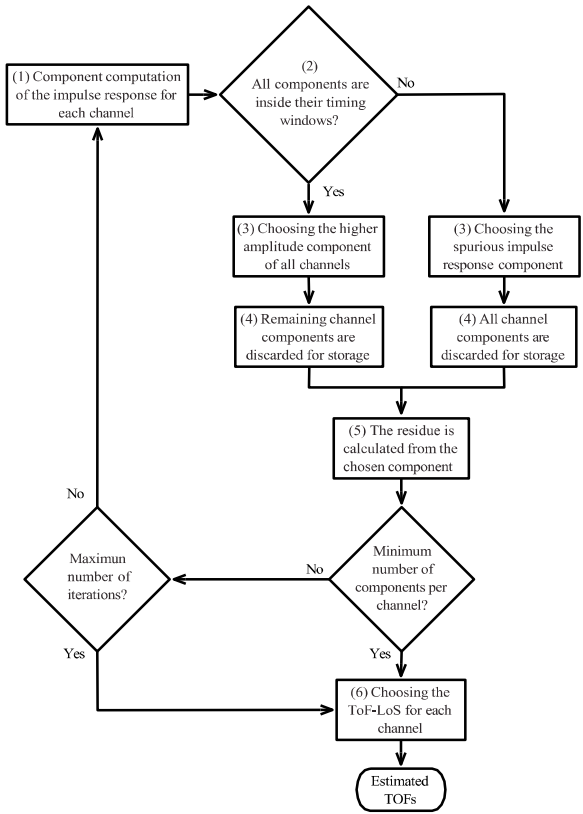

Figure 4. General block diagram of the proposed MCA.

## IV. Experimental Results

To evaluate the performance of the MCA, some experimental tests have been conducted in two different environments. The first test has been carried out in a small and empty box-shaped room (6.4×3.5×2.8 m³) in the Faculty of Science at the University of Extremadura (UEx). This room was chosen to favor the multipath appearance due to its reduced dimensions and its bad acoustic conditions since the walls, ceiling and floor are made of materials (plasterboard, plaster and laminate flooring) with high acoustic reflection coefficients (all of them $\geq 0.9$). The receiver is located at the origin of the $xy$ plane at 1 m high, that is, at coordinates $(0,0,1)$. Furthermore, to force multipath conditions a flat panel has been placed near the receiver in order to generate artificial echoes. Fig. 5 illustrates this room and the measurement equipment used to perform the experimental tests.

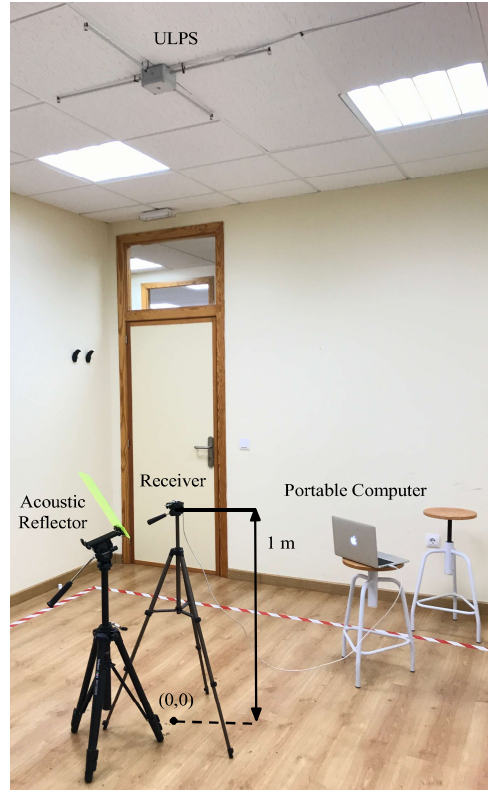

Figure 5. Experimental environment and measurement equipment used in UEx.

For the above mentioned test point $(0,0,1)$, a set of 250 measurements of the receiver position with and without the Acoustic Reflector (AR) were estimated. Such a number of measurements has been selected to evaluate statistically the error for the position estimates and to plot a suitable Cumulative Distribution Function (CDF) for experimental tests. Anyway, it is worth noting that a position estimate is obtained for every measurement. Running the proposed algorithms in a PC, the processing time required for an iteration (one measurement) is about 0.4 s. For the first set of measurements where the AR implies a strong multipath at the receiver, an study of the MCA performance according to the configuration of its design parameters $\gamma$ and $M$ has been conducted.

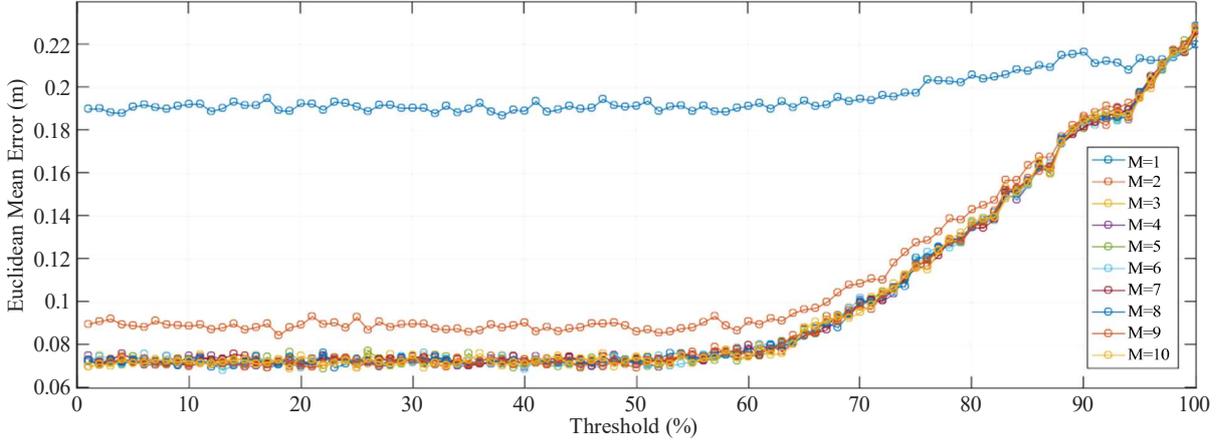

Figure 6. MCA performance depending on the design parameters $\gamma$ and $M$.

Fig. 6 represents the Euclidean Mean Error obtained for different combinations of the number $M$ of impulse response components per channel $i$ (from $M = 1$ to $M = 10$) with different detection thresholds $\gamma$ (from $\gamma = 1\%$ to $\gamma = 100\%$). From these results two main evidences can be concluded. The first one refers to the number $M$ of impulse response components per channel $i$ which are necessary to ensure a good MCA performance. As can be observed, from $M = 3$ to $M = 10$ coefficients per channel $i$ the MCA performance remains practically the same in terms of accuracy. For this reason, it seems logical to choose a minimum of $M = 3$ components per channel $i$ to avoid an unnecessary computational cost. On the other hand, it can be observed that the value of thresholds $\gamma$ above 50% of the amplitude of the highest impulse response component $\hat{h}^i$ degrades the MCA performance, since it increases the probability of failure when the LoS component is selected. Taking this into account, a low enough $\gamma = 10\%$ threshold is defined in order to achieve a proper selection of the LoS impulse response component.

After these parameters have been fixed at $M = 3$ and $\gamma = 10\%$, the Error Cumulative Distribution Function (ECDF) of this set of measurements with and without the MCA application is calculated. In addition, the maximum number of iterations for all channels $J$ was set at 32. Fig. 7 shows the ECDF obtained in each case.

As is shown in Fig. 7, the MCA implies a significant improvement in the positioning accuracy. In this sense, the application of the proposed algorithm improves the system performance, obtaining estimates of the receiver position with errors below 10 cm for 95% of measurements. On the contrary, when the classical method (based on single correlation and thresholding) is used to determine the ToAs, the errors in the receiver position estimation increase up to 44 cm for the same 95% of measurements.

A second set of measurements has been acquired where no acoustic reflector is used. These measurements were processed, not only to show the negative effects caused by the AR, but also to evaluate the MCA performance in the absence of multipath. Considering the results obtained using the AR, the improvement is clear, thus dropping from an error of 44 cm to 9 cm for the 95% of measurements. With regard to the results obtained by the classical method without multipath, these seem to be better (less than 2.5 cm) for the 50% of measurements; nevertheless, for the remaining 50% of measurements, the MCA provides similar accuracy with an extra 7% (from 90% to 97%) of measurements with errors below 8 cm. This deterioration in the classical method is motivated by the adverse acoustic conditions in the working environment. Fig. 8 depicts the resulting ECDF in each case.

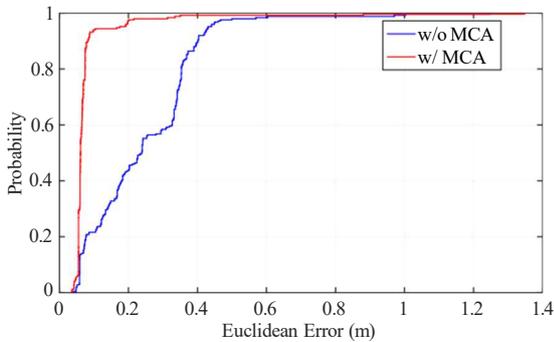

Figure 7. ECDF with AR and with and without the MCA application.

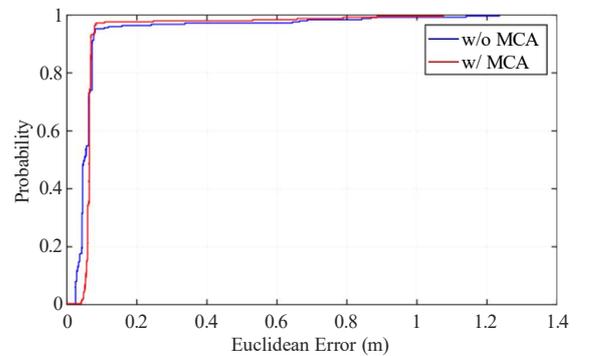

Figure 8. ECDF not involving the AR, with and without the MCA application.

In order to evaluate the proposed algorithm in a real test environment, an additional set of experiments has been conducted at the School of Engineering from the University of Alcalá (UAH). Fig. 9 a) shows the distribution of the test points in a complex environment (laboratory), that allows the behavior of the MCA to be verified. For every test point, a set of 100 measurements has been captured in order to obtain statistically the error and variance of the position estimates. Fig. 9 b) shows the average of those position estimates with and without applying the MCA.

In addition, the variance for every position estimate is plotted by an ellipse. As can be observed, only the point P21 presents some problems because it is surrounded by furniture and close to the wall. For the remaining points there is a significant improvement when the proposed MCA is applied, in comparison with the classical approach, since the estimated points are very close to the real ones and their variances are low with respect to the variances obtained without the MCA. Finally, Fig. 10 shows the ECDF of all distance errors for both approaches, the MCA and the classical one. The one proposed here provides a relevant improvement in general terms, more than 3 cm in the 80% of cases, with an error lower than 5.31 cm.

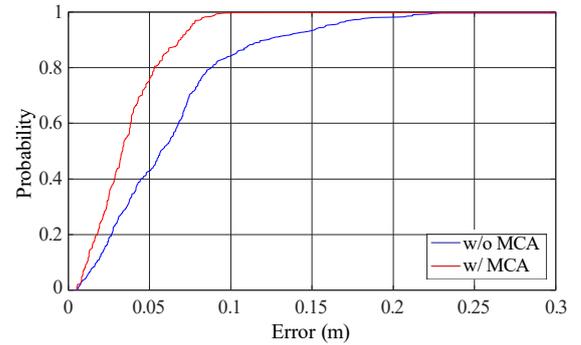

Figure 10. ECDF for every position considered at the UAH laboratory.

*Discussion*

To stablish a fair comparative, a table showing some key features of the proposed work and the most closely related ones has been included. As shown in Table I, the four acoustic positioning systems have high accuracies (below 10 cm) and operate in relative small positioning areas. On the one hand, the first experiment of the proposed work and [29] have the same coverage and both were conducted inside the same box-shaped room. This room is made of highly acoustic reflective materials which strengthen an intense multipath effect. Besides, both systems use a small number of beacons to reduce the computational tasks and improve the positioning rate. On the other hand, [28] proposes a beacon autocalibration system using audible signals which can be affected by multipath and [27] presents a local positioning system based on a similar algorithm to the one used in this work in order to compensate for possible LoS obstructions. These systems, [28] and [27], offer positioning with better accuracy and coverage than the proposed one at the expense of using a larger number of beacons and longer processing times.

## V. CONCLUSIONS

This work has presented a Multipath Compensation Algorithm (MCA) to enhance the performance of an Ultrasonic Positioning System under adverse multipath conditions, when the receiver is installed near walls or/and obstacles. The proposed algorithm is based on the accurate estimation of the environment impulse response from which the corresponding LoS for each channel is obtained.

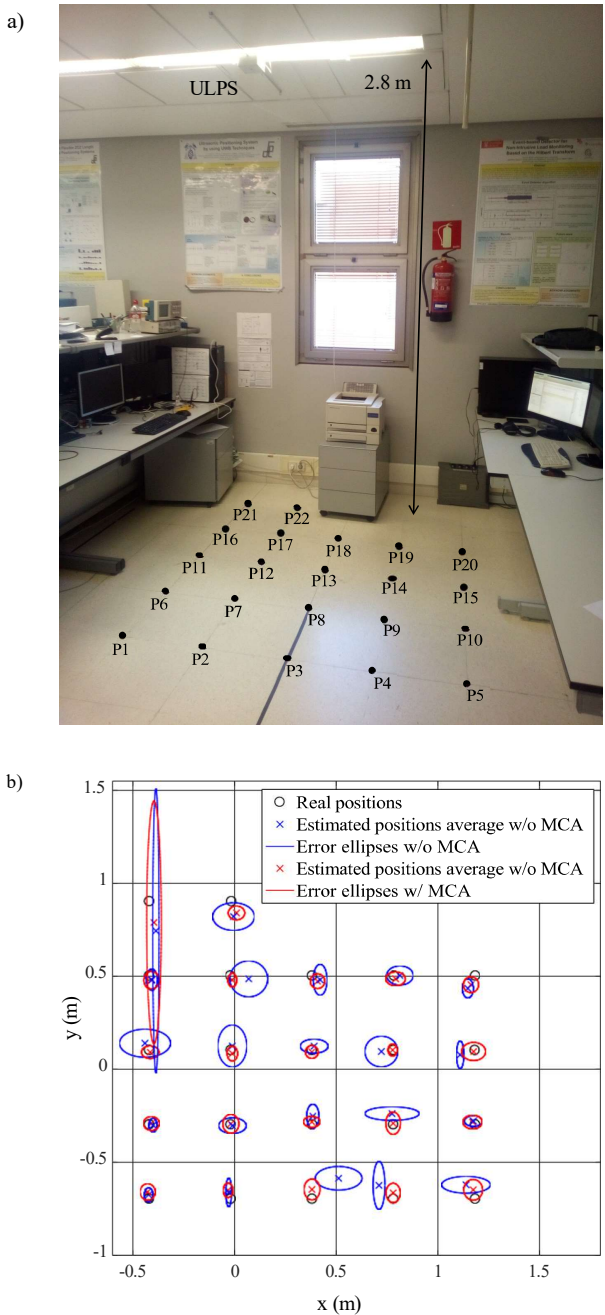

Figure 9. Environment and test points locations at the UAH laboratory (a); mean and variance for the position estimates (b).

Table I
COMPARATIVE TABLE WITH OTHER RELATED SYSTEMS.

| Work | Technology | Errors (cm) | Multipath Comp. | Coverage (m²) | Signals | $f_c$ (kHz) | Pos. rate (Hz) | # Beacons | Processing Unit |
|------|-----------|-------------|-----------------|---------------|---------|-------------|----------------|-----------|-----------------|
| Proposed | T-CDMA | < 9 for 95% of meas. | Yes | 3x3 | 255-bit Kasami codes | 41.67 | 2.5 | 5 | Laptop |
| [27] | TDMA | < 4.5 with LoS | No | 5.2x7.5 | Pseudo-Random Chip | Not Given | > 0.86 | 11 or 8 | Smartphones and Tablets |
| [28] | BeepBeep and MDS | MAE 1.10 | Yes | 6x6 | M-Sequences | 6 | 0.25 | 8 | Smartphones |
| [29] | CDMA | <10 for 90% of meas. | Yes | 3x3 | 63-bit Kasami codes | 16 | 1.5 | 4 | Tablet |

Experimental results in two different environments have been conducted in order to evaluate the performance of this proposal. The first set of experiments was carried out in an empty box-shaped room, where a severe multipath was artificially generated with a plane reflector. The second one took place in a complex environment such as a laboratory, where a grid of test points was arranged between furniture and obstacles, which generate multiple reflections and negatively affect the system performance. In both environments experimental results have confirmed the expected improvements, thus obtaining a significant reduction of the positioning errors. For the first environment, a study of the MCA performance depending on the choice of its design parameters $\gamma$ and $M$ has been conducted. The Error Cummulative Distribution Function (ECDF) has shown that positioning errors are reduced from 44 cm to 9 cm for the 95% of the measurements. With regard to the second environment, the ECDF has also shown that errors are reduced significantly: in this case from $8.78$ cm to $5.31$ cm for the 80% of the measurements.

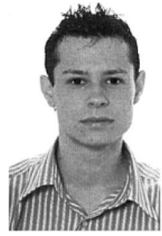

**David Gualda** received the B.S. degree in Electronics Systems and the M.Sc. degree in Advanced Electronics System, from the University of Alcalá (Spain), in 2009 and 2011, respectively; and the Ph.D. degree in Electronics in 2016. He is currently at the Department of Electronics of the University of Alcalá with a Postdoctoral Research Contract. His main research interests are in the areas of ultrasonic indoor location, signal processing and information fusion.

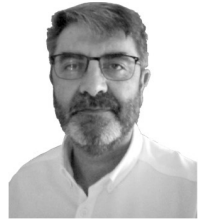

**José Manuel Villadangos** received the Ph.D. degree from the University of Alcalá, Alcalá de Henares, Spain, in 2013. He is currently an Associate Professor of Electronic Digital Systems with the Electronics Department,University of Alcalá. His current research interests include multisensor integration, electronic systems for mobile robots, digital and embedded systems, low-level ultrasonic signal processing, and local positioning systems

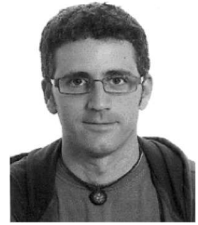

**Álvaro Hernández (M'06, SM'15)** received the Ph.D. degree from the University of Alcalá, Spain, and Blaise Pascal University, France, in 2003. He is currently an Associate Professor of Digital Systems and Electronic Design with the Electronics Department, University of Alcalá. His research areas are multisensor integration, electronic systems for mobile robots, and digital and embedded systems.

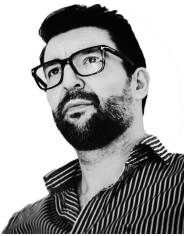

**Teodoro Aguilera** received his Physics degree in 2011 and his Ph.D. in Electronics in 2016 all of them from the University of Extremadura (Spain). He was an Assistant Professor of Automation for three years in the Department of Electrical Engineering, Electronics and Automation at the University of Extremadura. Currently he is a research member of the Sensory Systems Group of this university where his work lies in the design of Acoustic Local Positioning Systems (ALPS) based on mobile devices.

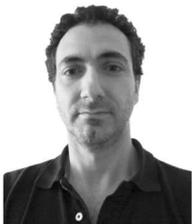

**Fernando J. Álvarez (M'07, SM'17)** received his Physics degree from the University of Sevilla (Spain), his Electronic Engineering degree from the University of Extremadura (Spain), his M.S. degree in Signal Theory and Communications from the University of Vigo (Spain), and his Ph.D. degree in Electronics from the University of Alcalá (Spain). During 2008 he was a postdoctoral 'José Castillejo' Fellow in the Intelligent Sensors Laboratory, Yale University (USA). He is currently an Associate Professor of Digital Electronics in the Department of Electrical Engineering, Electronics and Automation at the University of Extremadura, Spain, where he is also the head of the Sensory Systems Group. His research areas of interest include local positioning systems, acoustic signal processing and embedded computing.

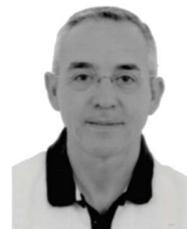

**Jesús Ureña (M'06, SM'15)** received the B.S. degree in Electronics Engineering and the M.S. degree in Telecommunications Engineering from Universidad Politéc'nica de Madrid, Madrid, Spain, in 1986 and 1992, respectively; and the Ph.D. degree in Telecommunications from the Universidad de Alcalá, Alcalá de Henares, Spain, in 1998. Since 1986, he has been with the Department of Electronics, University of Alcalá, currently as a Professor. He has collaborated in several educational and research projects in the area of electronic control and sensorial systems for mobile robots and wheelchairs and in the area of electronic distributed systems for railways. His current research interests are in the areas of ultrasonic signal processing, local positioning systems (LPSs), and sensory systems for railway safety.